\documentclass[prd,aps,tightenlines,superscriptaddress,amsmath,amssymb,amsfonts]{revtex4-2}
\usepackage{amssymb,amsmath,amsthm,amsbsy,epsfig,color,graphicx,times}
\usepackage[ansinew]{inputenc}
\usepackage[english]{babel}
\usepackage{color}
\usepackage{braket}
\usepackage{tabularx}
\usepackage{array}
\usepackage{mathtools}

\begin{document}

\title{A background independent notion of causality}

\author{A. Capolupo}
\email{capolupo@sa.infn.it}
\affiliation{Dipartimento di Fisica ``E.R. Caianiello'' Universit\`{a} di Salerno, and INFN -- Gruppo Collegato di Salerno, Via Giovanni Paolo II, 132, 84084 Fisciano (SA), Italy}

\author{A. Quaranta}
\email{anquaranta@unisa.it}
\affiliation{Dipartimento di Fisica ``E.R. Caianiello'' Universit\`{a} di Salerno, and INFN -- Gruppo Collegato di Salerno, Via Giovanni Paolo II, 132, 84084 Fisciano (SA), Italy}

\begin{abstract}

We develop a notion of causal order on a generic manifold as independent of the underlying differential and topological structure. We show that sufficiently regular causal orders can be recovered from a distinguished algebra of sets, which plays a role analogous to that of topologies and $\sigma$ algebras. We then discuss how a natural notion of measure can be associated to the algebra of causal sets.    

\end{abstract}

\maketitle

\section{Introduction}

Since the discovery of particle creation by black holes \cite{Hawking75} and the establishment of black hole thermodynamics \cite{BHT1,BHT2,BHT3}, a deep relation between the causal structure of spacetime and the tenets of thermodynamics has been unveiled \cite{Jacobson}. The causal structure is arguably  the core of the geometric side of general relativity, and it is known that any spacetime may be reconstructed, up to conformal transformations, from its causal order \cite{Malament}. This is enough to suspect that causality may be, in some sense, more fundamental than the differential character of the spacetime manifold. It is imaginable to construct theories of spacetime in which no differential structure is given, but only a causal order is assumed.  One may go further, and speculate that a future theory of quantum gravity may not have a natural differential manifold structure at all. For this reason a notion of causality which is abstracted and separated from topogical and differential notions might be necessary. The purpose of this work is to give a set theoretic construction of causal orders, based on a distinguished algebra of sets, which is in principle distinct from any topology given on the manifold. The construction is somewhat similar to that of topologies and $\sigma$-algebras, except that union (completion) are replaced by a new kind of operation, and there is an inherent built-in duality. The algebra of causal sets stems intuitively from the causal structure of Lorentzian manifolds and springs from the physical intuition of the double cone structure. It has the virtue, compared to other constructions such as Alexandroff topologies \cite{Alexandroff}, to be a direct reflection of the physical notion of causal structure as learned from our experience with Lorentzian manifolds. In addition, we shall find that a natural measure, inspired by statistical mechanics, may be introduced on the algebra of causal sets. The horizon entropy (and in particular the black hole entropy) will be seen to arise as a special case, on Lorentzian manifolds, of a more general construction. The paper is structured as follows: in section II we discuss the paradigmatic example of Minkowski spacetime, fix the notation, and determine the distinguished subsets that are causally relevant; in section III we abstract the notion of causal sets and show that the partial order relation can be reconstructed from the algebra; in section IV we introduce the notion of causal measure and formal entropy and analyze the specific case of horizon entropy; finally section V is devoted to the conclusions.

\section{Causal structure on Manifolds}

Let us start by reviewing the causal order in Minkowski spacetime $M^{1,3}$, that is $\mathbb{R}^4$ equipped with the Lorentzian metric 
\begin{equation}
 \eta_{\mu \nu} = \mathrm{diag} \left( 1,-1,-1,-1\right)  \ .
\end{equation}
Setting $A^2 = \eta_{\mu \nu}A^{\mu} A^{\nu}$, the vector $A^{\mu}$ is:
\begin{itemize}
 \item $A^2 > 0$ timelike
 \item $A^2 = 0$ lightlike
 \item $A^2 < 0$ spacelike \ .
\end{itemize}
The convention is the opposite if the $-,+.+.+$ signature is employed. A vector $A^{\mu}$ wich is either timelike or lightlike $A^2 \geq 0$ is said to be causal. The Lorentzian signature induces a pre-order on the manifold $\vartriangleleft$ as follows:
\begin{equation}\label{Preorder}
 x \vartriangleleft y \Leftrightarrow (x-y)^2 \geq 0
\end{equation}
for any two points $x,y \in M^{1,3}$. On a more general Lorentzian manifold, the condition is replaced by the existence of a causal curve with endpoints $x$ and $y$. The relation \eqref{Preorder} is reflexive and transitive. The introduction of the causal order requires another piece of information, the \textbf{time orientation}. For any point $x$, the set of points $y$ such that $x-y$ is causal, which we denote $C(x)$ has the structure of a double cone, i. e.
\begin{equation}
 C(x) = C_{1}(x) \cup C_{2}(x)
\end{equation}
with $C_1(x) \cap C_2(x) = \lbrace x \rbrace$. Time orientation is a smooth choice of one of the cones $C_i (x)$ at each $x$, the future directed cone $C_{+}(x)$ at $x$, such that for any $x,y,z \in M^{1,3}$
\begin{equation}\label{transitivity}
 z \in C_{+} (y) \  \&  \ y \in C_{+} (x) \Rightarrow z \in C_{+} (x) \ .
\end{equation}
we denote by $C_{-} (x)$ the remaining cone. There are a couple of warnings to make regarding the special character of Minkowski space as a vector space. On a general Lorentzian manifold these definitions still make sense provided that the notion of pre-order is modified as described above. The condition $x-y$ causal is to be replaced by the existence of a causal curve with endpoints $x$ and $y$. The sets $C_i(x)$ may then not have the shape of cones. The latter is retained locally on the tangent space at any point $x$, which, by definition, is isomorphic to $M^{1,3}$. The time orientation can be induced by the choice of a timelike vector field. The natural choice on Minkowski space is the one induced by the coordinates, whose timelike component we denote $x^0$. The future cone is 
\begin{equation}
 C_{+}(x) = \lbrace y \in M^{1,3} | (x-y)^2  \geq 0 \ \& \ y^0 \geq x^0 \rbrace \ . 
\end{equation}
With time orientation we can define the causal order as the partial order relation
\begin{equation}
 x \preceq y \Leftrightarrow y \in C_{+} (x)
\end{equation}
which is immediately seen to be reflexive, antisymmetric and transitive (due to eq. \eqref{transitivity}). There is an equivalent definition
\begin{equation}
 x \preceq y \Leftrightarrow C_{+} (y )\subseteq C_{+} (x) \ .
\end{equation}
We give the basic terminology
\newtheorem{Causality}{Definition}
\begin{Causality}
 A pair $(M,\preceq)$ with $\preceq$ a partial order on $M$ will be called equivalently a \textbf{poset}, a \textbf{causality} or a \textbf{causal order} on $M$.
\end{Causality}
Generally we will not make any assumption on the character of $M$ as a manifold. 

\subsection{A side on notation and terminology}

In a Lorentzian manifold there are at least two notions of partial order, the chronological order for relations that are exclusively timelike, and the causal order which includes lightlike separations. We will only be concerned with the causal order (although all the considerations still apply to the chronological order). Light or null cones are often referred \cite{Wald} to the distinguished subsets of the tangent space, rather than to the regions of the manifold. Here we will refer to the latter with this terminology. When the causal structure of Lorentzian Manifolds is concerned it is understood that the incomplete diamonds (see below) coincide with the causal past and the causal future of a given point $J^{\pm}(p)$.

\subsection{Reverse causality}

The double cone structure has evidently, built-in, a reverse causality. Consider a time orientation $\lbrace C_{+} (x), C_{-}(x) \rbrace$ and define the reverse orientation as the complementary choice at each $x$. This amounts to switching the future and past cones. If $\mathcal{O} = \left\lbrace \left(C_{+}(x) , C_{-} (x) \right) | x \in M^{1,3} \right)\rbrace$ is the set of pairs of cones, inversion of the time orientation can be regarded as a map
$
 T: \mathcal{O} \rightarrow \mathcal{O} 
$
such that
\begin{equation}
 T \left( \left(C_{+} (x), C_{-}(x) \right) \right) = \left(C_{-} (x), C_{+}(x) \right) \ .
\end{equation}
Clearly $T^2 = \mathrm{Id}$. The reverse causal order $\preceq^{-1}$ is the one induced by the reversed pairs, i.e. 
\begin{equation}
 x \preceq^{-1} y \Leftrightarrow C_{-} (y ) \subseteq C_{-} (x) \ .
\end{equation}
Obviously
\begin{equation}
 x \preceq^{-1} y \Leftrightarrow y \preceq x \ .
\end{equation}
If regarded as a map on the manifold $T: M^{1,3} \Rightarrow M^{1,3}$, $T$ is simply the time reversal (regarded as an active transformation on $M^{1,3}$) $T (x^0,x^j) = (-x^0,x^j)$. We have
\begin{equation}
 T(x) \preceq T(y) \Leftrightarrow y \preceq x
\end{equation}
so that we can identify 
\begin{equation}
 x \preceq^{-1} y \Leftrightarrow T(x) \preceq T (y) \ .
\end{equation}
From now on we shall regard $T:M \rightarrow M$ as an invertible automorphism that reverses the order. It is clear that the double cone structure carries information on both a causal order and its reverse.

\subsection{Abstracting the cone structure: $\nabla$ and $\Delta$ sets}

From the point of view of the causal order relation, the cones have some important properties. Let us focus on the future cones $C_{+}(x)$ for definiteness:
\begin{enumerate}
 \item \textbf{Completeness} For any $y \in C_{+} (x)$, the set $C[x,y] = \lbrace z \in M^{1,3} | x \preceq z \ \& z \preceq y \rbrace$ is contained in $C_{+} (x)$. This is trivial from the definition of $C_{+} (x)$ and essentially states that the cones contain all the possible causal curves extending from the basepoint $x$ to any of their elements $y \in C_{+}(x)$.
 \item \textbf{Divergence} For any $y,z \in C_{+} (x)$ that are not related by the causal order $\preceq$ there exists a third point $w \in C_{+} (x)$ such that $w \preceq y$ and $w \preceq z$. This is pretty obvious, since at least the basepoint $w=x$ satisfies the relations for each $y,z$. This means that no matter how $y,z$ are picked in $C_{+}(x)$, they always share a common ``past'' from some point on.
\end{enumerate}
Past cones are the exact mirror, they are causally complete and are \textbf{convergent} in the sense that
for any $y,z \in C_{-} (x)$ that are not related by the causal order $\preceq$ there exists a third point $w \in C_{+} (x)$ such that $y \preceq w$ and $z \preceq w$.
Let us formalize these definitions for later usage.

\newtheorem{Diamonds}{Definition}[section]
\begin{Diamonds}
 Given a causal order $\preceq$ on a manifold $M$ we define the \textbf{diamond} with endpoints $x,y$ as
\begin{equation}
 C[x,y] = \lbrace z \in M | x \preceq z \preceq y \rbrace \ .
\end{equation}
Notice that $C[x,y] \neq C[y,x]$ in general, and that a diamond may be the empty set. We define the \textbf{incomplete upper(lower) diamond} with endpoint $x$ as the sets
\begin{equation}
 C[\infty,x] (C[x,\infty]) = \lbrace z \in M | z \preceq x (x \preceq z) \rbrace
\end{equation}
\end{Diamonds}

\newtheorem{CauComplete}[Diamonds]{Definition}
\begin{CauComplete}
 A subset $U \subseteq M$ is \textbf{causally complete} if and only if 
\begin{equation}
 C[x,y] \subseteq U \ \forall \ x,y \in U \ .
 \end{equation}
\end{CauComplete}

\newtheorem{ConvSets}[Diamonds]{Definition}
\begin{ConvSets}
A subset $U \subseteq M$ is \textbf{convergent (divergent)} iff for any $x,y \in U$ not $\preceq$-related there exists $z \in U$ such that 
\begin{equation}
 x \preceq z \  \&  \ y \preceq z \ (z \preceq x \ \& \ z \preceq y)
\end{equation}
\end{ConvSets}
This definition is equivalent to that of a directed set in the direct and reverse causality respectively. 

\newtheorem{DeltaSets}[Diamonds]{Definition}
\begin{DeltaSets}
 A subset of $M$ is a $\Delta$ ($\nabla$) set iff it is causally complete and convergent (divergent).
\end{DeltaSets}

Notice that by the above definition, the collection of causal sets ($\Delta$ or $\nabla$) is much larger than the collection of cones. It indeed includes cones that are cut up to a given spacelike surface. A more detailed description of the causal sets on a Lorentzian Manifold shall be given in section 4.  

\newtheorem{CrossingProperty}[Diamonds]{Definition}
\begin{CrossingProperty}
 A manifold $M$ endowed with the partial order relation $\preceq$ has the \textbf{crossing property} iff for any $x,y,z,w \in M$ with $x,y \preceq z,w$ and $x,y$ not $\preceq$-related, at least one of the pairs $C[x,z] \ \& \ C[y,w]$ or  $C[x,w] \ \& \ C[y,z]$ has non-empty intersection.
\end{CrossingProperty}

\vspace{1 mm}
\emph{Example} The $(1+1)d$ Minkowski spacetime with the standard causal order trivially satisfies the crossing property. Let $p,q$ be any two points spacelike to each other (not $\preceq$-related) and let $r,s$ be any two distinct points satisfying $x,y \preceq r,s$. At least one of the two pairs of line segments $[x,r], [y,s]$ or $[x,s], [y,r]$ has an intersection at some point $p$. Since each of the line segments is timelike, and $r^0,s^0 \geq p^0 \geq x^0, y^0$, $p$ is necessarily in $C[x,r]$ and $C[y,s]$, by definition of the latter. 

We prove the following

\newtheorem{IntThe}[Diamonds]{Proposition}
\begin{IntThe}
 On a manifold satisfying the crossing property, the intersection of two $\Delta$ sets is $\Delta$, the intersection of two $\nabla$ sets is $\nabla$.
 \begin{proof}
 Let $A,B$ be $\Delta$ sets. For any $x,y \in A$, $C[x,y] \subseteq A$ and for any $x,y \in B$, $C[x,y] \subseteq B$. Then for any $x,y \in A \cap B$, we have $C[x,y] \subseteq A$ and $C[x,y] \subseteq B$, i. e. $C[x,y] \subseteq A \cap B$. This proves completeness. Convergence is a little trickier. Take any two $x,y$ not $\preceq$-related. Because $A,B$ are $\Delta$, there exist $r \in A$ and $s \in B$ such that $x,y \preceq r,s$. If $r=s$, we are done, therefore suppose that $r \neq s$. At least one of the following pairs has a nonempty intersection: $C[x,r] \ \& \ C[y,s]$ or  $C[x,s] \ \& \ C[y,r]$. Any of the points $p$ in the intersection satisfies $x,y \preceq p$. Moreover, since $A,B$ are both complete, any such point $p$ must be included in both, i.e. $p \in A \cap B$.
 \end{proof}
\end{IntThe}

\newtheorem{TimeRevIm}[Diamonds]{Proposition}
\begin{TimeRevIm}
 Let $T: M \rightarrow M$ be the order reversion automorphism. The image under $T$ of a $\Delta$ set is $\nabla$ and viceversa.
 \begin{proof}
  Let $A$ be a $\Delta$ set and let $x,y \in A$. Since $A$ is complete $C[x,y] \subseteq A$, and $T(C[x,y])\subseteq T(A)$. For any point $z \in C[T(y),T(x)]$ it holds that
\begin{equation}
 x \preceq T^{-1}(z) \preceq y
\end{equation}
i.e. $T^{-1}(z) \in C[x,y]$. Consequently $z = T(T^{-1}(z)) \in  T(C[x,y]) \subseteq T(A)$, so that $T(A)$ is complete. For any $x,y \in T(A)$ not $\preceq$-related there exists $z \in A$ such that $T^{-1}(x) \preceq z \ \& \ T^{-1}(y) \preceq z$. But then the point $T(z) \in T(A)$ satisfies $T(z) \preceq x,y$. This proves that $T(A)$ is divergent. The proof is similar for the converse statement.   
 \end{proof}
\label{TimeRevIm}
\end{TimeRevIm}

\newtheorem{TimeRevPro1}[Diamonds]{Proposition}
\begin{TimeRevPro1}
 Let $A,B$ be any two subsets of $M$. Then $T(A \cap B) = T(A) \cap T(B)$ and $T(A \cup B) = T(A) \cup T(B)$
 \begin{proof}
  Let $z \in A \cap B$. Since $z \in A$, $T(z) \in T(A)$ and since $z \in B$, $T(z) \in T(B)$. Therefore $T(A\cap B) \subseteq T(A) \cap T(B)$. Similarly, suppose $z \in T(A) \cap T(B)$. Given that $z \in T(A)$, $T^{-1}(z) \in A$, and considered that $z \in T(B)$, $T^{-1}(z) \in B$. But then $T^{-1}(z) \in A \cap B$ and thus $z=T(T^{-1}(z)) \in T( A\cap B)$, implying $T(A) \cap T(B) \subseteq T(A \cap B)$. The proof of the second statement is similar. Let $z \in A \cup B$. Because $z \in A$ or $z \in B$, $T(z) \in T(A)$ or $T(z) \in T(B)$, i.e. $z \in T(A) \cup T(B)$. Therefore $T(A \cup B) \subseteq T(A) \cup T(B)$. Then consider $z \in T(A) \cup T(B)$. Since $z \in T(A)$ or $z \in T(B)$, $T^{-1}(z) \in A$ or $T^{-1}(z) \in B$, i. e. $T^{-1}(z) \in A \cup B$. Therefore $z=T(T^{-1}(z)) \in T(A \cup B)$ and $T(A) \cup T(B) \subseteq T(A \cup B)$. 
 \end{proof}
\label{TimeRevPro}
\end{TimeRevPro1}
We remark that the statement holds for infinitely many intersections as long as we declare $T(\O{}) = \O{}$.

\newtheorem{Boundedness}[Diamonds]{Definition}
\begin{Boundedness}
 A $\Delta$ ($\nabla$) set $A$ is said to be \textbf{bounded} if it admits an upper (lower) bound $x \in A$, i. e.  such that for any $y \in A$,
\begin{equation}
 y \preceq x  \ \ \  (x \preceq y) \ .
 \end{equation}
\end{Boundedness}

\newtheorem{BCorr}[Diamonds]{Proposition}
\begin{BCorr}
 If a $\Delta$ ($\nabla$) set $A$ is bounded, there is a unique upper (lower) bound in  $A$.
 \begin{proof}
  The proof is trivial. Given $x,y \in A$ two upper bounds, $x \preceq y$ and $y \preceq x$, and this is only possible if $x=y$.
The proof is similar for $\nabla$ sets. 
 \end{proof}
\end{BCorr}
We call the unique upper (lower) bound $x \in A$ the upper (lower) \textbf{vertex} of $A$. It is quite clear from the definition that any bounded $\Delta$ ($\nabla$) set is a subset of an incomplete upper (lower) diamond. There exist sets that are both $\Delta$ and $\nabla$. Diamonds $C[x,y]$ are trivially so, and possess both an upper vertex $y$ and a lower vertex $x$. 

\newtheorem{StrictlyCausal}[Diamonds]{Definition}
\begin{StrictlyCausal}
 Denote by $C_{\Delta} (M) = \lbrace U \subseteq M | U \ \mathrm{is} \ \Delta \rbrace$ and $C_{\nabla} (M) = \lbrace U \subseteq M | U \ \mathrm{is} \ \nabla \rbrace$ the collections of $\Delta$ and $\nabla$ subsets of $M$. We say that a set $A$ is \textbf{strictly} $\Delta$ iff $A \in C_{\Delta}(M) - C_{\nabla}(M) \equiv C_{+}(M)$ and it is \textbf{strictly} $\nabla$ iff $A \in C_{\nabla}(M)-C_{\Delta}(M) \equiv C_{-} (M)$.
\end{StrictlyCausal}
Clearly $C_{+}(M)$ and $C_{-}(M)$ are disjoint by definition.

We have seen that $\Delta$ and $\nabla$ sets come with two natural algebraic operations
\begin{eqnarray*}
 \cap &:& C_{\Delta}(M) \times C_{\Delta}(M) \rightarrow C_{\Delta}(M) \\
 \cap &:& C_{\nabla}(M) \times C_{\nabla}(M) \rightarrow C_{\nabla}(M) \\
 T &:& C_{\Delta}(M) ( C_{\nabla}(M)) \rightarrow C_{\nabla}(M) (C_{\Delta} (M)).
\end{eqnarray*}
We can introduce another operation, the causal union, deemed to be for causal sets the equivalent of union for open sets in a topology. As the name suggests, it defines a larger causal set out of two given causal sets. 

\newtheorem{CausalUnion}[Diamonds]{Definition}
\begin{CausalUnion}
 The \textbf{causal union} of two $\Delta$ ($\nabla$) sets $A$ and $B$ is the smallest $\Delta$ ($\nabla$) set containing $A$ and $B$:
\begin{equation}
 A \cup_{C} B = \bigcap_{X \in C_{\Delta}(M) (C_{\nabla}(M)) : A \cup B \subseteq X } X \ .
\end{equation}
The causal union of two sets $A,B$ that are both $\Delta$ and $\nabla$ is the smallest set which is both $\Delta$ and $\nabla$ and contains $A \cup B$. The causal union of a set $A$ which is both $\Delta$ and $\nabla$ with a $\Delta$ ($\nabla$) set $B$ is the smallest $\Delta$ ($\nabla$) set containing $A \cup B$. The causal union of a strictly $\Delta$ and a strictly $\nabla$ set is the empty set. 
\end{CausalUnion}
The last two statements of the above definitin are introduced in order that the collection of $\Delta$ and $\nabla$ sets be closed under causal union. It is important to realize that the operation of causal union descends only from the causal order. In the next section, we will assume the causal union as given, and derive a partial order relation from it. The following properties hold

\newtheorem {CauUniPro}[Diamonds]{Proposition}
\begin{CauUniPro}
It is understood that the following properties hold for both convergent and divergent unions. For simplicity, we will assume the involved sets are all convergent (trivial sets included). Let $A,B,C \subseteq M$ be convergent causal sets. Then
\begin{enumerate}
\item $A,B \subseteq A \cup_{\mathcal{C}} B$
\item $ A \cup_{\mathcal{C}} A = A$
\item $ A \cup_{\mathcal{C}} ( B \cup_{\mathcal{C}} C) = (A \cup_{\mathcal{C}} B ) \cup_{\mathcal{C}} C$
\item $ C \cap ( A \cup_{\mathcal{C}} B) = (C \cap A) \cup_{\mathcal{C}} (C \cap B)$
\item $ A \cup_{\mathcal{C}} (B \cap C) = (A \cup_{\mathcal{C}} B) \cap ( A \cup_{\mathcal{C}} C)$
\item $T(A \cup_{\mathcal{C}} B) = T(A) \cup_{\mathcal{C}} T(B)$
\end{enumerate}
\begin{proof}
Properties I and II come trivially out of the definition. To facilitate the proof of the other properties, let us introduce the families of sets
\begin{equation}
\mathcal{C}(A,B) \doteq \lbrace V \subseteq M \ : \ V \ is \ \triangle \  and  \ A \cup B \subseteq V \rbrace
\end{equation}
and
\begin{equation}
\mathcal{C} (A,B,C) \doteq \lbrace V \subseteq M \ : \ V \ is \ \triangle \  and \ A \cup B \cup C \subseteq V \rbrace
\end{equation}
Then, by definition, one has
\begin{equation}
A \cup_{\mathcal{C}} ( B \cup_{\mathcal{C}} C) = \bigcap_{V \in \mathcal{C}(A, B \cup_{\mathcal{C}} C)} V
\end{equation}
Of course one has $ \mathcal{C}(A, B \cup_{\mathcal{C}} C) \subseteq \mathcal{C} (A,B,C)$, so that, defining
\begin{equation}
A \cup_{\mathcal{C}} B \cup_{\mathcal{C}} C \doteq \bigcap_{V \in \mathcal{C}(A,B,C)} V
\end{equation}
we have
\begin{equation}
A \cup_{\mathcal{C}} B \cup_{\mathcal{C}} C \subseteq A \cup_{\mathcal{C}} ( B \cup_{\mathcal{C}} C)
\end{equation}
and with a similar reasoning, one finds
\begin{equation}
A \cup_{\mathcal{C}} B \cup_{\mathcal{C}} C \subseteq (A \cup_{\mathcal{C}}  B) \cup_{\mathcal{C}} C
\end{equation}
so that
\begin{equation}
\left[(A \cup_{\mathcal{C}} B) \cup_{\mathcal{C}} C\right] \cap \left[A \cup_{\mathcal{C}} ( B \cup_{\mathcal{C}} C)\right] \subseteq A \cup_{\mathcal{C}} B \cup_{\mathcal{C}} C 
\end{equation}
On the other hand, it is clear that
\begin{equation}
A \cup_{\mathcal{C}} B \subseteq A \cup_{\mathcal{C}} B \cup_{\mathcal{C}} C
\end{equation}
\begin{equation}
B \cup_{\mathcal{C}} C \subseteq A \cup_{\mathcal{C}} B \cup_{\mathcal{C}} C
\end{equation}
\begin{equation}
A,C \subseteq A \cup_{\mathcal{C}} B \cup_{\mathcal{C}} C
\end{equation}
Thus
\begin{equation}
A \cup_{\mathcal{C}} B \cup_{\mathcal{C}} C \in \mathcal{C}(A \cup_{\mathcal{C}} B, C) 
\end{equation}
\begin{equation}
A \cup_{\mathcal{C}} B \cup_{\mathcal{C}} C \in \mathcal{C}(A , B \cup_{\mathcal{C}} C)
\end{equation}
and finally, because of the definition of causal union:
\begin{equation}
(A \cup_{\mathcal{C}} B) \cup_{\mathcal{C}} C = [(A \cup_{\mathcal{C}} B) \cup_{\mathcal{C}} C] \cap [A \cup_{\mathcal{C}} B \cup_{\mathcal{C}} C] = A \cup_{\mathcal{C}} B \cup_{\mathcal{C}} C 
\end{equation}
\begin{equation}
= [A \cup_{\mathcal{C}} B \cup_{\mathcal{C}} C] \cap [A \cup_{\mathcal{C}} (B \cup_{\mathcal{C}} C)] = A \cup_{\mathcal{C}} (B \cup_{\mathcal{C}} C)
\end{equation}
proving III. To prove IV, consider that, by definition,
\begin{equation}
C \cap (A \cup_{\mathcal{C}} B) = C \cap \bigcap_{V \in \mathcal{C}(A,B)}V = \bigcap_{V \in \mathcal{C}(A,B)} V \cap C
\end{equation}
Since any element of $\mathcal{C}(C \cap A, C \cap B)$ is of the form $U  \cap C$ for some $U \in \mathcal{C}(A,B)$, we can write
\begin{equation}
C \cap ( A \cup_{\mathcal{C}} B) = \bigcap_{V \in \mathcal{C}(C \cap A, C \cap B)}V = (C \cap A) \cup_{\mathcal{C}} (C \cap B)
\end{equation}
With a similar reasoning, we can prove V. Indeed, by definition 
\begin{equation}
A \cup_{\mathcal{C}} (B \cap C) = \bigcap_{V \in \mathcal{C}(A, B \cap C)} V
\end{equation} 
but since any set $V \in \mathcal{C}(A, B \cap C)$ contains the union $ A \cup (B \cap C) = (A \cup B) \cap (A \cup C)$, it is evidently of the form $V = X \cap Y$ for some $X \in \mathcal{C}(A,B)$ and $Y \in \mathcal{C}(A,C)$. Therefore we can write
\begin{equation}
A \cup_{\mathcal{C}} (B \cap C) = \bigcap_{X \in \mathcal{C}(A,B)} \bigcap_{Y \in \mathcal{C}(A,C)} X \cap Y = \bigcap_{X \in \mathcal{C}(A,B)} X \cap \bigcap_{Y \in \mathcal{C}(A,C)} Y = (A \cup_{\mathcal{C}} B) \cap (A \cup_{\mathcal{C}} C)
\end{equation}
The last property requires a little more effort. We first prove that $T(\mathcal{C}(A,B)) = \mathcal{C}(T(A),T(B))$. For each $z \in A \cup B$, $T(z) \in T(A \cup B) = T(A) \cup T(B)$ by virtue of proposition \ref{TimeRevPro}. Therefore each set containing $A \cup B$ is mapped by $T$ to a set containing $T(A \cup B) = T(A) \cup T(B)$, that is, $T(\mathcal{C}(A,B)) \subseteq \mathcal{C}(T(A),T(B))$. On the other hand, each set containing $T(A) \cup T(B) = T(A \cup B)$ is mapped by $T^{-1}$ to a set containing $A \cup B$. Therefore $T^{-1}(\mathcal{C}(T(A),T(B))) \subseteq \mathcal{C}(A,B)$ and $\mathcal{C}(T(A),T(B)) = T (T^{-1}(\mathcal{C}(T(A),T(B)))) \subseteq T(\mathcal{C}(A,B))$.

Now, by definition
\begin{equation*}
 T(A \cup_{\mathcal{C}} B) = T \left(\bigcap_{V \in \mathcal{C}(A,B)} V\right) = \bigcap_{V \in \mathcal{C}(A,B)} T(V)
\end{equation*}
where the last equality follows from proposition \ref{TimeRevPro}. Finally, due to the statement just proven we can write
\begin{eqnarray*}
 T(A \cup_{\mathcal{C}} B) =\bigcap_{U \in T(\mathcal{C}(A,B))} U  =\bigcap_{U \in \mathcal{C}(T(A),T(B))} U  = T(A) \cup_{\mathcal{C}} T(B)
\end{eqnarray*}
with the last equality following from the definition.
\end{proof}
\label{CauUniPro}\end{CauUniPro}

With the introduction of the causal union, each of the collections $C_{\Delta}(M)$ and $C_{\nabla}(M)$ acquires an algebraic structure analogous to topologies and $\sigma$-algebras. Since the intersection of an infinite number of sets is at most the empty set, and the causal union of an infinite number of sets cannot exceed the whole manifold $M$, $C_{\Delta}(M)$ and $C_{\nabla}(M)$ are closed under countably many intersection and countably many causal unions if $\O{}$ and $M$ belong to $C_{\Delta}(M)$ and $C_{\nabla}(M)$. Only the empty set belongs to $C_{\Delta}(M), C_{\nabla}(M)$ by definition (as the empty set satisfies the relevant axioms trivially), while $M$ need not belong to either of the collections. In general only closure under countably many intersections and finitely many causal union is ensured.  

\newtheorem{TimeRevInv}[Diamonds]{Proposition}
\begin{TimeRevInv}
 The image of a strictly $\Delta$-set under $T$ is strictly $\nabla$ and the image of a strictly $\nabla$-set under $T$ is strictly $\Delta$.
 \begin{proof}
  The proof follows trivially from \ref{TimeRevInv}, for suppose that $A= T(B)$ is both $\Delta$ and $\nabla$,  then $B$ is also $\Delta$ and $\nabla$.
 \end{proof}

 \label{TimeRevInv}
\end{TimeRevInv}

\section{The abstract algebra of causal sets}

Up to now we have assumed an underlying order relation $\preceq$ on the base space $M$ and determined two special class of subsets of $M$ with respect to $\preceq$. We now go the other way around, assigning, abstractly, the data corresponding to $\Delta$ and $\nabla$ sets, without assuming any partial order on $M$. We give the following

\newtheorem{AbstractAlgebra}{Definition}[section]
\begin{AbstractAlgebra}
 An \textbf{algebra of causal sets} over a set $M$ is a fifthuple $(M,C_{\Delta}(M),C_{\nabla}(M),\cup_{\mathcal{C}},T)$ where
 \begin{enumerate}
  \item $C_{\Delta}(M)$ and $C_{\nabla}(M)$ are collections of subsets of $M$ such that $\O{} \in C_{\Delta}(M)$,$\O{} \in C_{\nabla}(M)$ and singletons $\lbrace p \rbrace$ for $p \in M$ belong to both, and which are closed under (countably many) intersections and (finitely many) causal unions
  \begin{eqnarray*}
  && \cap : C_{\Delta}(M) \times C_{\Delta}(M) \rightarrow C_{\Delta}(M) \ \ , \ \ \ \ \ \ \ \ \cup_{\mathcal{C}} : C_{\Delta}(M) \times C_{\Delta}(M) \rightarrow C_{\Delta}(M) \\
  && \cap : C_{\nabla}(M) \times C_{\nabla}(M) \rightarrow C_{\nabla}(M) \ \ , \ \ \ \ \ \ \ \ \cup_{\mathcal{C}} : C_{\nabla}(M) \times C_{\nabla}(M) \rightarrow C_{\nabla}(M)
  \end{eqnarray*}
  \item $T$ is an invertible map on $C_{\Delta}(M) \cup C_{\nabla}(M)$such that $T(C_{\Delta}(M))= C_{\nabla}(M)$, $T(C_{\nabla}(M)) = C_{\Delta}(M)$ and $T(\O{}) = \O{}$. Moreover $T$ satisfies the properties \ref{TimeRevPro} (it commutes with intersection and union). 
  \item The causal union operation $\cup_{\mathcal{C}}$, together with reversion $T$, satisfies the properties of \ref{CauUniPro}. 

 \end{enumerate}

\end{AbstractAlgebra}
In the following we prove that an algebra of causal sets contains at least as much information as a causality on $M$. We shall explicitly construct the partial order relation from the fifthuple given above. Let us start by providing some auxiliary definitions. As before we denote the collections of strictly causal sets as $C_{+} (M) = C_{\Delta}(M) - C_{\nabla}(M)$ and $C_{-}(M) = C_{\nabla}(M)-C(\Delta)(M)$. We also denote $C_{\pm}(p) = \lbrace A \in C_{\pm}(M) | p \in A \rbrace$.

\newtheorem{Ribboned}[AbstractAlgebra]{Definition}
\begin{Ribboned}
 A pair of strictly $\Delta$ and $\nabla$ sets passing through $p \in M$, $(A,B) \in C_{+}(p) \times C_{-}(p)$ is \textbf{ribboned} over p iff 
 \begin{equation*}
  A \cap B = \lbrace p \rbrace \ .
 \end{equation*}
 We call the set $R(p) = \lbrace (A,B)  \in C_{+}(p) \times C_{-}(p) | (A,B) \ \mathrm{is} \ \mathrm{ribboned} \ \mathrm{over} \ p \rbrace$ the \textbf{ribbon} over $p$.
\end{Ribboned}
Notice that the ribbon may be empty, and that the trivial pair $\left(\lbrace p \rbrace, \lbrace p \rbrace \right)$ is not in $R(p)$, since by definition singletons are both $\Delta$ and $\nabla$. The ribbon may also be trivial in a certain sense, as we clarify below.
\newtheorem{Density}[AbstractAlgebra]{Definition}
\begin{Density}
 A pair $(A,B) \in R(p)$ is \textbf{dense} in $R(p)$ if and only if for any non trivial (not the singleton $\lbrace p \rbrace$) subsets of the form $A' = A \cap V$ and $B' = B \cap W$ for some $V,W \in C_{+}(p) \cup C_{-}(p)$, there exist non-trivial $A'' \subseteq A'$ and $B'' \subseteq B'$ such that $(A'',B'') \in R(p)$.

\end{Density}
\newtheorem{Regularity}[AbstractAlgebra]{Definition}
\begin{Regularity}
 A point $p$ has a \textbf{regular ribbon} iff
 \begin{enumerate}
  \item Each $(A,B) \in R(p)$ is dense in $R(p)$
  \item For any two pairs $(A,B), (C,D) \in R(p)$ such that
  $
   (A \cup C) \cap  (B \cup D) = \lbrace p \rbrace 
  $
 it holds that 
 \begin{equation*}
  (A \cup_{\mathcal{C}} C) \cap  (B \cup_{\mathcal{C}} D) = \lbrace p \rbrace
 \end{equation*}

 \end{enumerate}

\end{Regularity}
The conditions for a regular ribbon ensure that there are enough pairs to construct a partial order relation. We simply will not put in relation points that do not have a regular ribbon. Their importance will become apparent in a moment. For points with a regular ribbon we define the following relation

\newtheorem{Congruence}[AbstractAlgebra]{Definition}
\begin{Congruence}
 Two pairs $(A,B),(C,D) \in R(p)$ are \textbf{congruent}, denoted $(A,B) \simeq (C,D)$ iff 
  $(A \cup_{\mathcal{C}} C,B \cup_{\mathcal{C}} D) \in R (p)$. 
\end{Congruence}
 
 \newtheorem{ConPro}[AbstractAlgebra]{Lemma}
 \begin{ConPro}
  If $(A,B) \simeq (C,D)$, it holds that $A \cap D = B \cap C = \lbrace p \rbrace$. 
  \begin{proof}
   By definition of congruence
   \begin{equation*}
    (A \cup_{\mathcal{C}} C) \cap (B \cup_{\mathcal{C}} D) = \lbrace p \rbrace
   \end{equation*}
 By repeated use of the properties of \ref{CauUniPro} we can write
 \begin{equation*}
  (A \cup_{\mathcal{C}}C) \cap ( B \cup_{\mathcal{C}} D) = \left[(A\cup_{\mathcal{C}}C) \cap B \right] \cup_{\mathcal{C}} \left[(A\cup_{\mathcal{C}}C) \cap D \right]= (A \cap B) \cup_{\mathcal{C}} (C\cap B) \cup_{\mathcal{C}} (A \cap D) \cup_{\mathcal{C}} (C \cap D) = \lbrace p \rbrace
 \end{equation*}
The last equality holds only if each of the intersections contains no more than $p$. On the other hand, since all the sets involved contain $p$, one has $A \cap B = C \cap B = A \cap D = C \cap D = \lbrace p \rbrace$.
  \end{proof}
\label{ConPro}
 \end{ConPro}

 \newtheorem{EquiRelCon}[AbstractAlgebra]{Theorem}
 \begin{EquiRelCon}
  Congruence is an equivalence relation on regular ribbons.
  \begin{proof}
   Reflexivity and symmetry are obvious, because $A \cup_{\mathcal{C}} A = A$ and $A \cup_{\mathcal{C}} B = B \cup_{\mathcal{C}} A$ for any $A,B$. To prove transitivity, suppose that $(A,B) \simeq (C,D)$ and $(C,D) \simeq (E,F)$. Consider the set
   \begin{equation*}
    (A \cup E) \cap (B \cup F) = (A \cap B) \cup (A \cap F) \cup (E \cap B) \cup (E \cup F)
   \end{equation*}
 where the equality follows from elementary set theoretic identities. Since $(A,B) \in R(p)$ and $(E,F) \in R(p)$, $A \cap B = \lbrace p \rbrace = E \cap F$. Considered that all the sets involved contain $p$ as an element, we can write
 \begin{equation*}
    (A \cup E) \cap (B \cup F) =  (A \cap F) \cup (E \cap B) \ .
   \end{equation*}
   Now, suppose that $A \cap F$ contains points other than $p$. Given that $p$ is regular, there exists a non-trivial subset $G \subseteq A \cap F \subseteq A$ that belongs to $C_{+}(p)$. Considered that $(C,D)$ is dense in $R(p)$, at least one of the following statements is true
   \begin{eqnarray*}
&&\lbrace p \rbrace \subset G \cap C \subseteq A \cap F \cap C \\ && \lbrace p \rbrace \subset G \cap D \subseteq A \cap F \cap D
   \end{eqnarray*}
But $(A,B) \simeq (C,D)$, implying that $A \cap D = \lbrace p \rbrace$ and $(C,D) \simeq (E,F)$, so that $C \cap F = \lbrace p \rbrace$ by the lemma \ref{ConPro}. Then each of the intersections on the right hand side equals $\lbrace p \rbrace$, leading to the contradiction $\lbrace p \rbrace \subset \lbrace p \rbrace$. We conclude that $A \cap F = \lbrace p \rbrace$. By a similar reasoning it is shown that $E \cap B = \lbrace p \rbrace$. Overall, we have proven that
\begin{equation*}
 (A \cup E) \cap (B \cup F) = \lbrace p \rbrace \ .
\end{equation*}
Finally, since $p$ has a regular ribbon by hypotesis, it follows that $(A \cup_{\mathcal{C}} E, B \cup_{\mathcal{C}} F) \in R(p) $.
  \end{proof}

 \end{EquiRelCon}

\newtheorem{ConCla1}[AbstractAlgebra]{Lemma}
\begin{ConCla1}
 Let $R(p)$ be a regular ribbon and $(A,B), (C,D) \in R(p)$ any two pairs. Then at least one of the following statements holds
 \begin{eqnarray*}
  A \cap C = \lbrace p \rbrace \\
  B \cap C = \lbrace p \rbrace
 \end{eqnarray*}
and similar for $D$
 \begin{eqnarray*}
  && A \cap D = \lbrace p \rbrace \\
  && B \cap D = \lbrace p \rbrace \ .
 \end{eqnarray*}
 \begin{proof}
  Suppose that both $A \cap C$ and $B \cap C$ contain points other than $p$. Then $A \cap C$ and $B \cap C$ are non trivial subsets of $A$ and $B$, and by regularity there exist $G \subseteq A \cap C$ and $G \subseteq B \cap C$ such that $(G,H) \in R (p)$. Then $(A,B) \simeq (G,H)$, because $A \cup_{\mathcal{C}} G = A$ and $B \cup_{\mathcal{C}} H = H$; $(G,D),(G,B) \in R(p)$, because $G \subseteq C$ and $G \subseteq A$, and $(G,D) \simeq (G,B)$, because
  \begin{equation*}
   G \cap ( D \cup B) = (G \cap D) \cup (G \cap B) = \lbrace p \rbrace \cup \lbrace p \rbrace = \lbrace p \rbrace
  \end{equation*}
where the second equality stems a fortiori from $A \cap B = \lbrace p \rbrace = C \cap D$. Moreover $(A,B) \simeq (G,B)$ and $(G,D) \simeq (C,D)$ trivially, since $A\cup_{\mathcal{C}} G= A$ and $C\cup_{\mathcal{C}} G = C$. The following chain of congruences holds
\begin{equation*}
 (A,B) \simeq (G,B) \simeq (G,D) \simeq (C,D)
\end{equation*}
so that $(A,B) \simeq (C,D)$. Therefore, by the lemma \ref{ConPro}, $B \cap C = \lbrace p \rbrace$ and , which is a contradiction of the hypothesis. The proof is analogous for the $D$ statements. 
 \end{proof}
\label{ConCla1}
\end{ConCla1}

\newtheorem{ClassCon}[AbstractAlgebra]{Theorem}
\begin{ClassCon}
 For a regular ribbon $R(p)$ there exist at most two distinct classes of congruence.
 \begin{proof}
  Consider any three non-trival pairs $(A,B), (C,D), (E,F) \in R(p)$ and assume that no two of them are congruent. From lemma \ref{ConCla1}, without loss of generality, assume that
  \begin{equation}
   A \cap E = B \cap F = C \cap E = D \cap F = A \cap C = B \cap D = \lbrace p \rbrace
  \end{equation}
but then
\begin{equation*}
 (A \cup_{\mathcal{C}}C) \cap ( B \cup_{\mathcal{C}} D) = \left[(A\cup_{\mathcal{C}}C) \cap B \right] \cup_{\mathcal{C}} \left[(A\cup_{\mathcal{C}}C) \cap D \right]= (A \cap B) \cup_{\mathcal{C}} (C\cap B) \cup_{\mathcal{C}} (A \cap D) \cup_{\mathcal{C}} (C \cap D) = (C\cap B) \cup_{\mathcal{C}} (A \cap D)
\end{equation*}
Now, either at least one between $(C \cap B)$ and $(A \cap D)$ is non-trivial, or the intersection equals $\lbrace p \rbrace$, so that $(A,B) \simeq (C,D)$, against the hypothesis. Suppose that $(C \cap B)$ is non-trivial. Considered that $(A,B)$ is dense, there exists a non-trivial $G \subseteq C \cap B \in C_{+}(p)$  such that $(G,D) \in R(p)$, because$G \cup_{\mathcal{C}} C = C$. Then, since also $(E,F)$ is dense, at least one of the following statements is true
\begin{eqnarray*}
&& \lbrace p \rbrace \subset G \cap E \\
&& \lbrace p \rbrace \subset G \cap F \ .
\end{eqnarray*}
In the first case we have $\lbrace p \rbrace \subset G \cap E \subseteq (C \cap B) \cap E = \lbrace p \rbrace$, and in the second $\lbrace p \rbrace \subset G \cap F \subseteq (C \cap B) \cap F = \lbrace p \rbrace$, which are both contradictory. A similar conclusion stems from the non-triviality of $A \cap D$, so that the hypothesis must be ruled out. The proof is analogous for all the other possible combinations of statements from the lemma \ref{ConCla1}.
 \end{proof}
\label{ClassCon}
\end{ClassCon}

The above theorem ensures that every (non empty) regular ribbon admits either a single congruence class or two distinct classes.  

\subsection{Causal order from the algebra of sets}

Armed with the previous results we now proceed to construct a partial order relation from the algebra of causal sets. 

\newtheorem{PoRel}[AbstractAlgebra]{Definition}
\begin{PoRel}
 Let $p,q \in M$ have regular ribbons $R(p),R(q)$. We say that $p$ and $q$ are related if and only if there exist two (non-empty) classes of congruence $\alpha \in R(p)/\simeq$ and $\gamma \in R(q)/\simeq$ such that one of the following statement holds
 \begin{enumerate}
  \item $(A \cup_{\mathcal{C}} C, D) \in \gamma$ and $(A, B \cup_{\mathcal{C}} D) \in \alpha$
  \item $(A \cup_{\mathcal{C}} C, B) \in \alpha$ and $(C, B \cup_{\mathcal{C}} D) \in \gamma$
 \end{enumerate}

 $\forall (A,B) \in \alpha$ and $\forall (C,D) \in \gamma$. 
 \label{PoRel}
\end{PoRel}
The definition \ref{PoRel} only specifies when two points with regular ribbon are related. It is important to realize that the algebra of causal sets carries information on both a causal order and its reverse, so that it is a matter of convention to tell which is which. We adopot the following convention: we set $p \preceq q$ iff case $1$ of the definition \ref{PoRel} holds:

\newtheorem{SRel}[AbstractAlgebra]{Definition}
\begin{SRel}
 We declare $p \preceq q$  if and only if there exist two classes of congruence $\alpha \in R(p)/\simeq$ and $\gamma \in R(q)/\simeq$ such that 
$(A \cup_{\mathcal{C}} C, D) \in \gamma$ and $(A, B \cup_{\mathcal{C}} D) \in \alpha$
 $\forall (A,B) \in \alpha$ and $\forall (C,D) \in \gamma$. 
 \label{SRel}
\end{SRel}

\newtheorem{RRel}[AbstractAlgebra]{Theorem}
\begin{RRel}
 The relation defined in \ref{SRel} is a partial order relation among the points of $M$ with regular ribbon.
 \begin{proof}
  \begin{itemize}
   \item \emph{Reflexivity}. Let $p \in M$ and $R(p)$ regular. Given a congruence class $\alpha \in R(p)/\simeq$, it holds that for any $(A,B), (C,D) \in \alpha$
   $
    (A \cup_{\mathcal{C}} C , B \cup_{\mathcal{C}} D) \in \alpha 
   $, by definition. Given that $(A \cup_{\mathcal{C}} C, D)$ and  $(A \cup_{\mathcal{C}} C, B \cup_{\mathcal{C}} D)$ satisfy 
   \begin{equation*}
    \left[(A \cup_{\mathcal{C}} C) \cup_{\mathcal{C}} (A \cup_{\mathcal{C}} C)\right] \cap \left[(B \cup_{\mathcal{C}} D) \cup_{\mathcal{C}} D\right] = (A \cup_{\mathcal{C}} C) \cap B \cup_{\mathcal{C}} D) = \lbrace p \rbrace
   \end{equation*}
   we have that $(A \cup_{\mathcal{C}} C, D)$ is an element of $R(p)$ congruent to $(A \cup_{\mathcal{C}} C , B \cup_{\mathcal{C}} D)$, and a fortiori, congruent to $(A,B)$ and $(C,D)$. Then $(A \cup_{\mathcal{C}} C, D) \in \alpha$. Similarly $(A, B \cup_{\mathcal{C}} D) \in \alpha$. Therefore the relation \ref{SRel} holds for $\gamma = \alpha$.
   
   \item \emph{Antisymmetry}. It is obvious that, given the simmetry of the definition \ref{PoRel}, if the first statement corresponds to $p \preceq q$, the second corresponds to $q \preceq p$. To prove antisymmetry it is sufficient to demonstrate that the two statements can hold simultaneously if and only if $p = q$. Let $R(p)$ and $R(q)$ regular, and suppose that both the statements hold. Then $\forall (A,B) \in \alpha$ and $\forall (C,D) \in \gamma$ we have $(A,B \cup_{\mathcal{C}} D) \in \alpha$ and $(A \cup_{\mathcal{C}} C, B) \in \alpha$. But then, given the definition of congruence, $(A \cup_{\mathcal{C}} C, B \cup_{\mathcal{C}} D) \in R(p)$, implying a fortiori
   \begin{equation*}
    (A \cup C) \cap (B \cup D) = \lbrace p \rbrace \ .
   \end{equation*}
   On the other hand 
   \begin{equation*}
    (A \cup C) \cap (B \cup D) = (A \cap B) \cup (A \cap D) \cup (C \cap B) \cup (C \cap D) = \lbrace p \rbrace \cup (A \cap D) \cup (C \cap B) \cup \lbrace q \rbrace
   \end{equation*}
    where the last step follows because $(A,B) \in R(p)$ and $(C,D) \in R(q)$ by hypothesis. The two equations are consistent if and only if $p = q$.
   \item \emph{Transitivity} Let $p,q,r$ with regular ribbons $R(p), R(q), R(r)$ and suppose $p \preceq q$ and $q \preceq r$. Denote $\alpha \in R(p)/\simeq$, $\gamma \in R(q)/\simeq$ and $\epsilon \in R(r)/\simeq$ for which the relevant statements hold. For all $(A,B) \in \alpha, (C,D) \in \gamma, (E,F) \in \epsilon$, one has
   \begin{itemize}
    \item $(A \cup_{\mathcal{C}} C,D) \in \gamma$ and $(A, B \cup_{\mathcal{C}} D) \in \alpha$
    \item $(C \cup_{\mathcal{C}} E,F) \in \epsilon$ and $(C, D \cup_{\mathcal{C}} F) \in \gamma$
   \end{itemize}
   Picking $C' = A \cup_{\mathcal{C}} C$, with $(C',D) \in \gamma$, one has
   \begin{equation*}
    (C' \cup_{\mathcal{C}} E, F) = ( A \cup_{\mathcal{C}} C\cup_{\mathcal{C}} E, F) \in \epsilon
   \end{equation*}
Considered that $r \in A \cup_{\mathcal{C}} E$, and that, a fortiori $(A \cup_{\mathcal{C}} E) \cap F = \lbrace r \rbrace$, we conclude that $(A \cup_{\mathcal{C}} E, F) \in \epsilon$, because it is trivially congruent to $( A \cup_{\mathcal{C}} C\cup_{\mathcal{C}} E, F)$. Similarly, picking $D' = D \cup_{\mathcal{C}} F)$ we have
\begin{equation*}
 (A. B \cup_{\mathcal{C}} D') = (A, B \cup_{\mathcal{C}} D \cup_{\mathcal{C}} F) \in \alpha
\end{equation*}
which by the same reasoning above yields $(A, B \cup_{\mathcal{C}} F) \in \alpha$. We have managed to prove that $\forall (A,B) \in \alpha$ and $\forall (E,F) \in \epsilon$
$
 (A \cup_{\mathcal{C}} E, F) \in \epsilon
$
and $(A, B \cup_{\mathcal{C}} F) \in \alpha$, i. e. $p \preceq r$.

  \end{itemize}

 \end{proof}

\end{RRel}

We now have to check that the reversal $T$ acts as expected by reverting the order defined in \ref{SRel}. We have the following

\newtheorem{TimeRev1}[AbstractAlgebra]{Lemma}
\begin{TimeRev1}
 If $p \in M$ has a regular ribbon $R(p)$, the point $T(p) \in M$ has also a regular ribbon $R(T(p))$.
 \begin{proof}
  We prove this by explicit construction. Set $R(T(p)) = \lbrace (T(B),T(A)) | (A,B) \in R(p) \rbrace$. Since $p \in A,B$, evidently $T(p) \in T(A),T(B)$. Moreover, by the properties \ref{TimeRevPro}, $T(A) \cap T(B) = T(A \cap B) = \lbrace T(p)\rbrace$. Then $(T(B),T(A) \in C_{+}(T(p)) \times C_{-} (T(p))$, because of \ref{TimeRevInv}, and is evidently ribboned over $T(p)$. All of the possible ribboned pairs over $T(p)$ are of this form, since, suppose that $(C,D)$ is ribboned over $T(p)$. Then $(T^{-1}(D), T^{-1}(C))$ is ribboned over $p$, by the same arguments as above. Therefore $R(T(p))$ is indeed the ribbon over $T(p)$. Regularity follows immediately from the regularity of $R(p)$, because $T$ commutes with intersections, unions and causal unions.  
 \end{proof}
\label{TimeRev1}
\end{TimeRev1}

\newtheorem{TimeRev2}[AbstractAlgebra]{Theorem}
\begin{TimeRev2}
 If $p \preceq q$ according to \ref{SRel}, then $T(q) \preceq T(p)$. 
 \begin{proof}Let $\alpha \in R(p)/\simeq$ and $\gamma \in R(q)/\simeq$ be the two relevant congruence classes. Set $T( \alpha) = \lbrace (T(B), T(A)) | (A,B) \in \alpha\rbrace$ and $T( \gamma) = \lbrace (T(D), T(C)) | (C,D) \in \gamma\rbrace$. That $T(\alpha)$ and $T(\gamma)$ are congruence classes in $R(T(p))$ and $R(T(q))$, follows from \ref{TimeRev1}. By hypothesis $\forall (A,B) \in \alpha$ and $\forall (C,D) \in \gamma$ we have $(A \cup_{\mathcal{C}} C, D) \in \gamma$ and $(A, B \cup_{\mathcal{C}} D) \in \alpha$. But
 \begin{eqnarray*}
  && (T(D),T(A \cup_{\mathcal{C}} C)) = (T(D), T(A) \cup_{\mathcal{C}} T(C)) \in T(\gamma) \\
  && (T(B \cup_{\mathcal{C}} D), T(A)) = (T(B) \cup_{\mathcal{C}} T(D), T(A)) \in T(\alpha)
 \end{eqnarray*}
Therefore there exist classes of congruence $T(\gamma)$ and $T(\alpha)$ such that for any $(E,F) \in T(\gamma) $ and for any $(G,H) \in T(\alpha)$ it holds $(E \cup_{\mathcal{C}} G, H) \in T(\alpha)$ and $(E, F \cup_{\mathcal{C}} H) \in T(\gamma)$, which, by comparison with \ref{SRel}, implies $T(q) \preceq T(p)$.
 
  \end{proof}

\end{TimeRev2}
We conclude the section by discussing how the abstract definition \ref{SRel} coincides with the assigned causal order, when the latter is given and sufficiently regular. The identification proceeds as follows. When the causal order $\preceq$ is given we can explicitly construct a congruence class in the ribbon over $p$ by considering strictly $\nabla$ and $\Delta$ sets bounded respectively below and above by $p$. If the order is sufficiently regular (in the sense that causal sets are sufficiently dense), the causal union of sets $A,B$ with the same vertex $p$ is again a set with vertex $p$, because the causal union selects the smallest possible set containing $A$ and $B$. This can be immediately seen for the light cones in Minkowski spacetime. Now if $p \preceq q$, any $\Delta$ set bounded by $p$ can be enlarged to a $\Delta$ set bounded by $q$ and any $\nabla$ set bounded by $q$ can be enlarged to a $\nabla$ set bounded by $p$. This is the practical significance of the abstract definition \ref{SRel}.  The minimal requirements for the identification are the following (here all the operations are those induced by the assigned causal order $\preceq$):
\newtheorem{RegularCausality}[AbstractAlgebra]{Definition}
\begin{RegularCausality}
\begin{itemize}
 \item $\forall p \in M$, and for all $A,B \in C_{+}(p)$ bounded by $p$, $A \cup_{\mathcal{C}} B \in C_{+}(p)$ and is bounded by $p$.
 \item $\forall p \in M$, and for all $A,B \in C_{-}(p)$ bounded by $p$, $A \cup_{\mathcal{C}} B \in C_{-}(p)$ and is bounded by $p$
 \item The order is sufficiently dense that if $p \preceq q$ for each $A \in C_{+}(p)$ bounded by $p$, there exists $B \in C_{+}(q)$, bounded by $q$, such that $A \subseteq B$, and, similarly, for each $A \in C_{-}(q)$ bounded by $q$, there exists $B \in C_{-}(p)$ bounded by $p$, such that $A \subseteq B$.
\end{itemize}
We say that a causality $(M , \preceq)$ is \textbf{regular} if it satisfies the above requirements and the crossing property.
\end{RegularCausality}
The base manifold $M$ may fail to meet these requirements in some regions, in which case the domain of the identification is restricted to the subset of $M$ satisfying them. Given these considerations, the causal order induced by the abstract algebra of causal sets is equivalent to a regular causality. 

\section{Causal measure}

There is a meaningful way by which can weigh the elements of an algebra of causal sets. To understand this point, let us focus on the $C_{\nabla}(M)$ part, of the $\nabla$ sets of a given algebra. Let us recall the notion of number of microstates in statical mechanics, which for a system in equilibrium with total energy $E$ we denote $\Omega (E)$. The total number of microstates for a system made of two isolated subsystems, at energy $E_1$ and $E_2$ is $\Omega_{1+2} (E_1 + E_2) = \Omega_1(E_1) \Omega_2 (E_2) $, with $\Omega_i (E)$ the number of microstates for subsystem $i$. We want a multiplicative measure that mirrors, for causal sets, this property.

Let $A \in C_{\nabla}(M)$ and assign a function $\sigma: C_{\nabla}(M) \rightarrow [1,\infty)$. If $A = \lbrace p \rbrace$ consists of a single point, we assign $\sigma (A) = 1$.  We expect that if $A,B \in C_{\nabla}(M)$ are disjoint, we wish that $\sigma (A \cup B) = \sigma (A) \sigma (B) $. If $A,B$ are not disjoint, we have to remove the overcounting of the overlap $A \cap B$, so that we set $\sigma(A \cup B) = \frac{\sigma(A) \sigma(B)}{\sigma (A \cap B)}$. The two requirements are consistent if we additionally declare $\sigma (\O{}) = 1$. Yet we have to recall that the relevant operation for causal sets is not the union, but the causal union. Therefore we impose the milder requirement that $\sigma(A \cup_{\mathcal{C}} B) \geq \frac{\sigma(A) \sigma(B)}{\sigma (A \cap B)}$, with equality holding if $A \cup B= A \cup_{\mathcal{C}} B$. Then we have the following

\newtheorem{CausalMeasure}{Definition}[section]
\begin{CausalMeasure}\label{CMeasure}
 A (divergent) \textbf{causal measure} is a map $\sigma : C_{\nabla} (M) \rightarrow [1,\infty]$ such that
 \begin{enumerate}
  \item $\sigma(\O{}) = \sigma(\lbrace p \rbrace) = 1 \ \forall p \in M$
  \item $\forall A,B \in C_{\nabla}(M)$ it holds
  \begin{equation*}
   \sigma(A \cup_{\mathcal{C}} B) \geq \frac{\sigma(A) \sigma(B)}{\sigma (A \cap B)}
  \end{equation*}
  and equality is verified for $A \cup B = A \cup_{\mathcal{C}} B$.
 \end{enumerate}
\end{CausalMeasure}

The definition can be extended to all the subsets of $M$ in two different ways. For any $A \subseteq M$, let $C_{\nabla} (A) = \lbrace X \in C_{\nabla} (M)| A \subseteq X \rbrace$ and $C^*_{\nabla}(A) = \lbrace X \in C_{\nabla}(M) | X \subseteq A \rbrace$. We then define $\sigma (A) =\inf_{X \in C_{\nabla} (A)} \sigma (X)$ and $\sigma^*(A) = \sup_{X \in C*_{\nabla} (A)}\sigma (X)$. The two extensions need not coincide, and we will prefer $\sigma^*$ over $\sigma$. 

Similarly it is possible to define a convergent causal measure on $C_{\Delta}(M)$. A causal measure enjoys the following intuitive property:
\newtheorem{CausalMeasureProperties}[CausalMeasure]{Proposition}
\begin{CausalMeasureProperties}\label{CMeasure2}
If $A \subseteq B$, $\sigma (A) \leq \sigma(B)$. 
\begin{proof}
 If $A$ is the empty set or a singleton, the proof is trivial because $\sigma (A) = 1$ and $\sigma (B) \geq 1$ by definition. Suppose that $\sigma(A) > 1$. Write $B = (B-A) \cup (B \cap A)$, so that clearly $B = (B-A) \cup_{\mathcal{C}} (B \cap A)$.  Then 
 \begin{equation*}
  \sigma(B) = \frac{\sigma (B-A) \sigma(B \cap A)}{\sigma((B-A)\cap(B \cap A))} = \frac{\sigma(B-A) \sigma(A)}{\sigma(\O{})} = \sigma(B-A) \sigma(A)
 \end{equation*}
here, if $B-A$ is not a $\nabla$ sets, any of the two extensions $\sigma, \sigma^*$ is understood. Given that $\sigma(B-A) \geq 1$ by definition, we obtain the statement. 
\end{proof}

\end{CausalMeasureProperties}
For a given causal measure we can define the related \textbf{formal entropy} as $S(A)= k_B \log (\sigma(A))$, where $k_B$ is the Boltzmann constant. By virtue of the above proposition, $S(A)$ is additive in the sense that if $A \subseteq B$
\begin{equation*}
 S(B) = S(B-A) + S(A) \ .
\end{equation*}
This result extends by induction to any partition of a set $B$ by disjoint subsets $B= \bigcup_{i} X_i$, so that $S(B)= \sum_i S(X_i)$. Then the formal entropy has the properties of a measure in the Lebesgue sense, since it is also non-negative and satisfies $S(\O{})=0$. Of course we can go the other way around, assigning a formal entropy $S$ and defining the causal measure by $\sigma = \exp{S/k_B}$. 
We notice that the definition \ref{CMeasure} and the ensuing property \ref{CMeasure2} imply a certain character of monotonicity of the formal entropy, i. e. for $A \subseteq B$, $S(A) \leq S(B)$. This seemingly trivial aspect rules out some more exotic definitions of entropy, such as Tsallis-like entropies with $q>1$ (see for instance \cite{Tsallis2009}) for which the relation 
\begin{equation}
S_q(A \cup B) =S_q (A) + S_q(B) + \frac{(1-q)}{k_B}S_q (A) S_q(B)   
\end{equation}
would imply, for $A \subset B$, that the entropy of $B$
\begin{equation}
 S_q (B) = S_q \left(A \cup \left(B-A \right)\right) = S_q(A) + S_q (B-A) + \frac{(1-q)}{k_B}S_q (A) S_q(B-A)   
\end{equation}
which can be lesser than $S_q (A)$ for $q>1$. In a certain sense, this kind of exotic entropies goes against our naive intuition of causality.

\subsection{Horizon Entropy}

On Lorentzian manifolds there exists a natural notion of formal entropy. Let us focus, for simplicity, on Minkowski spacetime and characterize its causal sets. Pick a set of rectangular coordinates and denote them, for $p \in M^{1,3}$, $x_p^{\mu}\equiv (x^0_p,x^1_p,x^2_p,x^3_p)$. Naturally the causal order is equally described in any coordinate system related by proper Lorentz transformations, since the latter cannot affect causal relationships. We have seen that the diamonds $C[p,q]$ are both $\nabla$ and $\Delta$. Let $x^0_p$ and $x^0_q$ be the time coordinates and let $M(t_1,t_2) = \lbrace p \in M^{1,3} | t_1 \leq x^0_p \leq t_2  \rbrace$. We are essentially foliating $M^{1,3}$ by equal time surfaces and selecting the subset of $M$ comprised between the surfaces at $t_1$ and $t_2$. The sets $X[t,q] = M(t,x^0_q) \cap C[p,q]$ and $Y[p,t]= M(x^0_p,t) \cap C[p,q]$ for $x^0_p \leq t \leq x^0_q$ are respectively $\Delta$ and $\nabla$. Convergence and divergence are obvious, because $z \preceq q$ and $p \preceq z$  $\forall z \in C[p,q]$. To prove completeness, notice that $C[p,q] = X \cup Y$ and consider any two points $r,s \in X$. If $r,s$ are not related or $r=s$, $C[r,s] = \O{} (\lbrace r \rbrace)$, so there is nothing to prove. Then assume, without loss of generality, that $r \prec s$. Since $r,s \in X$, $t \leq x^0_r < x^0_s \leq x^0_q$, and all the points of $C[r,s]$ must have time coordinate between $t$ and $x^0_q$. Then $C[r,s] \subseteq M(t,x^0_q)$. On the other hand, by completeness of $C[p,q]$, $C[r,s] \subseteq C[p,q]$. Therefore $C[r,s] \subseteq M(t,x^0_q) \cap C[p,q] = X $. The proof is analogous for the completeness of $Y$. Other causal sets can be obtained by cutting the incomplete diamonds $C[p,\infty]$ and $C[\infty,p]$ up  to a time $t \leq (\geq) x^0_p$. Recall that incomplete diamonds themselves are $\Delta$ ($\nabla$) sets.   

Each of the causal sets has evidently a boundary made of a spacelike surface $S$ and a null region $H$. $S$ may not exist for sets that extend to infinity, but $H$ is always defined. We call $H$ the horizon of a given causal set and denote by $H_t = H \cap \Sigma_{t}$ the intersection of $H$ with the equal time surface at $t$. We define a formal entropy as follows:

\newtheorem{HorizonEnt}[CausalMeasure]{Definition}
\begin{HorizonEnt}
 Let $A$ be a $\nabla$ set and $H(A)$ its horizon (the null part of its boundary). If $H_t(A)$ is the intersection of $H(A)$ with the equal time surface at $t$, we define the \textbf{horizon entropy}
 \begin{equation*}
  S(A) = \alpha \sup_{t \in \mathbb{R}} \mathcal{A} (H_t(A))
 \end{equation*}
where $\mathcal{A}$ denotes the area of the $2d$ region $H_t(A)$ and $\alpha$ is a dimensionful positive constant.
\end{HorizonEnt}
The definition is naturally extended additively to disjoint unions $A \cup B$ by $S(A \cup B) =\alpha \sup_{t \in \mathbb{R}}  \mathcal{A} (H_t(A)) +  \alpha\sup_{t \in \mathbb{R}} \mathcal{A}(H_t(B))$. To understand what such a definition amounts to, let us compute the horizon entropy of a divergent set of the form $Y[p,t] = M(x^0_p,t) \cap C[p,\infty]$. The intersection $H_{t'}(Y[p,t])$ is the set of points $z \in M^{1,3}$ with time coordinate $t'$ such that $\eta_{\mu \nu}(x^{\mu}_z- x^{\mu}_p)((x^{\nu}_z- x^{\nu}_p)) = 0$. Written explicitly
\begin{equation*}
 (t'-x^0_p)^2 - (\pmb{z}-\pmb{p})^2 = 0 \Rightarrow R^2(t') = (\pmb{z}-\pmb{p})^2
\end{equation*}
where  $R^2(t') =(t'-x^0_p)^2$ and boldface letters are shorthand for the spatial components. This is just the $2$ sphere centered at $\pmb{p}$ with radius $R(t')$, so that its area is $\mathcal{A}(H_{t'}(Y[p,t])) = 4 \pi R^2 (t')$. This is evidently maximal when $t'=t$, (for $t' > t$ the intersection is empty) so that $S(Y[p,t]) = 4 \alpha \pi R^2(t)$. The horizon entropy takes upon a suggestive form if $p$ is chosen as the origin:
\begin{equation*}
 S(Y[0,t]) = 4 \alpha \pi t^2 \ .
\end{equation*}
If $S$ is to have units of entropy, $\alpha \sim \frac{k_B}{L^2}$ with $k_B$ the Boltzmann constant, $L$ a length, and a possible numerical coefficient. The Planck length $L=l_p = \sqrt{\frac{\hbar G}{c^3}}$ seems the natural choice. Bekenstein-Hawking entropy \cite{BHT1} results from $\alpha = \frac{k_B}{4l_p^2}$. Horizon entropy can be extended to any non-causal set $B$ by declaring $S(B) = \sup_{A \in C_{\nabla}(M) | A \subseteq B} S(A)$.
In Minkowski spacetime the horizon entropy is somewhat artificial, since it is not related to any form of curvature. Its definition in a general Lorentzian manifold is more significant, and only requires the replacement of equal $t$ surfaces with generic Cauchy surfaces. Consider, e.g. the Schwarzschild black hole $\mathcal{B}$ (see for instance \cite{Wald}). We can clearly fit in $\mathcal{B}$ a number of $\nabla$ sets that share the same horizon as $\mathcal{B}$ (but no larger) at a given Cauchy surface, so that the horizon entropy defined above shall coincide with the Bekenstein-Hawking entropy by construction. That the horizon entropy is a natural construction is also apparent from the celebrated result by Jacobson \cite{Jacobson}. 

\section{Conclusions}

We have provided the construction of the causal order relation from a distinguished algebra of sets (the causal sets), which assumes no topological or differential notions a priori. The construction is fully background independent, in that no geometry has to be assumed on the underlying set. This may turn out to be an important advantage in situations where no underlying geometry is given, as presumably it is the case in quantum gravity. We have discussed how causal sets admit a natural measure, which in the specific case of Lorentzian manifolds may be taken to coincide with the horizon entropy. In doing so, the relation between thermodynamics and the causal structure of spacetime has been made transparent. The construction presented here may be refined in future developments, and eventually be compared to other set algebraic constructions (topologies, $\sigma$-algebras, etc.) on more general grounds. In addition, it is worth noting that we have dealt with an exquisitely classical notion of causality. This may serve as the basis for the construction of a quantum notion of causality as a proper generalization.    

\section*{Acknowledgements}
Partial financial support from MIUR and INFN is acknowledged.
A.C. and G.L. also  acknowledge the COST Action CA1511 Cosmology
and Astrophysics Network for Theoretical Advances and Training Actions (CANTATA).

\end{document}